\documentclass[sigconf]{acmart}

\usepackage{booktabs} 
\usepackage{enumitem}
\usepackage{booktabs} 
\usepackage{graphicx}
\usepackage{subcaption}
\usepackage[singlelinecheck=false]{caption}
\usepackage{tabularx}
\usepackage{comment} 
\usepackage{amsmath}
\numberwithin{figure}{section}
\usepackage{lipsum} 
\usepackage{hyperref}
\usepackage{color}
\usepackage{multirow}
\setlist{nosep}

\begin{document}
\fancyhead{}
\settopmatter{printacmref=false}

\title{RELink: A Research Framework and Test Collection for Entity-Relationship Retrieval}

\author{Pedro Saleiro}
\affiliation{\institution{LIACC, FEUP, Universidade do Porto}}
\email{pssc@fe.up.pt}

\author{Nata\v{s}a Mili\'{c}-Frayling}
\affiliation{\institution{School of Computer Science, University of Nottingham}}
\email{natasa.milic-frayling@nottingham.ac.uk}

\author{Eduarda Mendes Rodrigues}
\affiliation{\institution{FEUP, Universidade do Porto}}
\email{eduarda@fe.up.pt}

\author{Carlos Soares}
\affiliation{\institution{INESC TEC, FEUP, Universidade do Porto}}
\email{csoares@fe.up.pt}

\renewcommand{\shortauthors}{B. Trovato et al.}

\begin{abstract}
Improvements of entity-relationship (E-R) search techniques have been hampered by a lack of test collections, particularly for complex queries involving multiple entities and relationships. In this paper we describe a method for generating E-R test queries to support comprehensive E-R search experiments. Queries and relevance judgments are created from content that exists in a tabular form where columns represent entity types and the table structure implies one or more relationships among the entities. Editorial work involves creating natural language queries based on relationships represented by the entries in the table. 
We have publicly released the RELink test collection comprising 600 queries and relevance judgments obtained from a sample of Wikipedia List-of-lists-of-lists tables. The latter comprise tuples of entities that are extracted from columns and labelled by corresponding entity types and relationships they represent. In order to facilitate research in complex E-R retrieval, we have created and released as open source the RELink Framework that includes Apache Lucene indexing and search specifically tailored to E-R retrieval. RELink includes entity and relationship indexing based on the ClueWeb-09-B Web collection with FACC1 text span annotations linked to Wikipedia entities. With ready to use search resources and a comprehensive test collection, we support community in pursuing E-R research at scale.

\end{abstract}

\copyrightyear{2017} 
\acmYear{2017} 
\setcopyright{acmcopyright}
\acmConference{SIGIR '17}{August 07-11, 2017}{Shinjuku, Tokyo, Japan}\acmPrice{15.00}\acmDOI{http://dx.doi.org/10.1145/3077136.3080756}
\acmISBN{978-1-4503-5022-8/17/08}

%
%




\maketitle
\section{Introduction}
In recent years, we have seen increased interest in using online information sources to find concise and precise information about specific issues, events, and entities \cite{yahya2016relationship}. For example, in response to the query: ``Low emission cars produced in the United States manufacturing plants'', one would expect an answer in terms of tuples $<$\textit{car brand}, \textit{US plant location or name}$>$ or  $<$\textit{car brand}, \textit{US plant location or name}, \textit{emission level}$>$. Since the Web offers an abundance of information, there are concerted efforts to extract entities and entity relationships from free text and to optimize entity-relationship (E-R) search.

The extraction process typically involves text processing using natural language processing (NLP) and machine learning methods to identify entity instances of a given type. Introducing new entity types and automating the extraction process requires manual effort to label the textual data and train the classifiers to identify correct instances in the text.  

Extracted entities and relationships are typically stored in a database or knowledge base.  Thus, one can leverage structured search to retrieve them in response to a user's query \cite{conrad1994system,yahya2016relationship}. However, pre-defining and constraining entity and relationship types reduce the range of queries that can be answered and therefore limit the usefulness of entity search, particularly when one wants to leverage free-text corpora such as the Web.  
Improvement of methods for both extraction and search is hampered by a lack of query sets and relevance judgments, i.e., golden standards that could be used to compare effectiveness of different methods. In this paper we introduce:
\begin{enumerate}
\item A low-effort semi-automatic method for acquiring instances of entities and entity relationships from tabular data. 
\item RELink Query Collection (QC) of 600 E-R queries with corresponding relevance judgments  
\item RELink Framework with resources that enable experimentation with multi-relationship E-R queries.
\end{enumerate}

Essential to our approach is the observation that tabular data typically includes entity types as columns and entity instances as rows. The table structure implies a relationship among table columns and enables us to create E-R queries that are answered by the entity tuples across columns.
Following this approach, we prepared and released the RELink QC comprising 600 E-R queries and relevance judgments based on a sample of Wikipedia \textit{List-of-lists-of-lists} tables. Furthermore, we used the ClueWeb-09-B Web collection with FACC1 text span annotations linked to Wikipedia entities to show how RELink can be used for E-R retrieval over Web content. We support E-R search through Apache Lucene indexing and search, tailored to multi-relationship entity retrieval. 

The query collection and the research framework are publicly available\footnote{\url{https://sigirelink.github.io/RELink/}}, enabling the community to expand the RELink Framework with additional document collections and alternative indexing and search methods. It is important to maintain and enhance the  RELink QC by providing updates to the existing entity types and creating new queries and relevant instances from additional tabular data.  

In the following sections we provide a rationale for our approach and situate the RELink QC and Framework in the context of related efforts. We provide a detailed account of the methods used to create the RELink collection and resources that are made available to support E-R search experiments within the RELink Framework.

\section{Related Collections}

To the best of our knowledge there are only two test collections specifically created for E-R retrieval: ERQ \cite{li2012entity} and COMPLEX \cite{yahya2016relationship}. Both support research retrieval of entities and relations from knowledge bases and neither provide complete relevance judgments. Consequently, researchers need to evaluate manually the answers they produce in their experiments. 

ERQ consists of 28 queries that were adapted from INEX17 and OWN28 \cite{li2012entity} initiatives. Twenty two of the queries express relationships, but already have one entity instance named and fixed in the query (e.g. \textit{``Find Eagles songs''}). Only 6 queries ask for pairs of unknown entities, such as ``\textit{Find films starring Robert De Niro and please tell directors of these films.}''. 

COMPLEX collection was created semi-automatically\cite{yahya2016relationship} and comprises relationship-centric queries for unknown entities, such as ``\textit{Currency of the country whose president is James Mancham}'', ``\textit{Kings of the city which led the Peloponnesian League.}''  and ``\textit{Who starred in a movie directed by Hal Ashby?}''. Among 70 queries, 60 involve entity pairs and 10 refer to entity triples.

\section{RELink Query Collection}

\subsection{Tabular Data and Entity Relationships}
Information that satisfies complex E-R queries is likely to involve instances of entities and their relationships dispersed across Web documents. Sometimes such information is collected and published within a single document, such as a Wikipedia page. In such cases, traditional search engines can provide excellent search results without applying special E-R techniques or considering entity and relationship types. Indeed, the data collection, aggregation, and tabularization has been done by a Wikipedia editor. 

That also means that a tabular Wikipedia content, comprising various entities, can be considered as representing a specific information need, i.e., the need that motivated editors to create the page in the first place. Such content can, in fact, satisfy many different information needs. We focus on exploiting tabular data for exhaustive search for pre-specified E-R types. In order to specify E-R queries, we can use column headings as entity types. All the column entries are then relevance judgments for the entity query. Similarly, for a given pair of columns that correspond to distinct entities, we formulate the implied relationship. For example the pair $<$car, manufacturing plant$>$ could refer to ``is made in'' or ``is manufactured in'' relationships. The instances of entity pairs in the table then serve as evidence for the specific relationship. This can be generalized to more complex information needs that involve multiple entity types and relationships.

Automated creation of E-R queries from tabular content is an interesting research problem. For now we asked human editors to provide natural language and structured E-R queries for specific entity types. Once we collect sufficient amounts of data from human editors we will be able to automate the query creation process with machine learning techniques. For the RELink QC we compiled a set of 600 queries with E-R relevance judgments from Wikipedia lists about 9 topic areas.   

\subsection{Selection of Tables}
Wikipedia contains a dynamic index ``\textit{The Lists of lists of lists}''\footnote{ \url{http://en.wikipedia.org/wiki/List_of_lists_of_lists}} which represents the root of a tree that spans curated lists of entities in various domains. We used a Wikipedia snapshot from October 2016 to traverse ``\textit{The Lists of lists of lists}'' tree starting from the root page and following every hyperlink of type ``\textit{List of}'' and their children. This resulted in a collection of 95,569 list pages. While most of the pages contain tabular data, only 18,903 include tables with consistent column and row structure. As in \cite{bhagavatula2013methods}, we restrict content extraction to \textit{wikitable} HTML class that typically denotes data tables in Wikipedia. We ignore other types of tables such as infoboxes. 

In this first instance, we focus on \textit{relational tables}, i.e., the tables that have a key column, referring to the \textit{main} entity in the table  \cite{lehmberg2016large}. For instance, the ''\textit{List of books about skepticism}'' contains a table ``\textit{Books}'' with columns ``Author'', ``Category'' and ``Title'', among others. In this case, the key column is ``Title'' which contains titles of books about skepticism. We require that any relationship specified for the entity types in the table must contain the  ``Title'' type, i.e., involve the ``Title'' column. 

In order to detect key columns we created a Table Parser that uses the set of heuristics adopted by Lehmberg et al. \cite{lehmberg2016large}, e.g., the ratio of unique cells in the column or text length. Once the key column is identified, the parser creates entity pairs consisting of the key column and one other column in the table. The content of the column cells then constitutes the set of relevant judgments for the relationship specified by the pair of entities. 

For the sake of simplicity we consider only those Wikipedia lists that contain a single relational table. Furthermore, our goal is to create queries that have verifiable entity and entity pair instances. Therefore, we selected only those relational tables for which the key column and at least one more column have cell content linked to Wikipedia articles.  

With these requirements, we collected 1795 tables. In the final step, we selected 600 tables by performing stratified sampling across semantic domains covered by Wikipedia lists. For each new table, we calcuated the Jaccard similarity scores between the title of the corresponding Wikipedia page and the titles of pages associated with tables already in the pool. By setting the maximum similarity threshold to 0.7 we obtained a set of 600 tables. 

\begin{figure}[t]
    \centering
        \centering
        \includegraphics[width= 1.0\columnwidth]{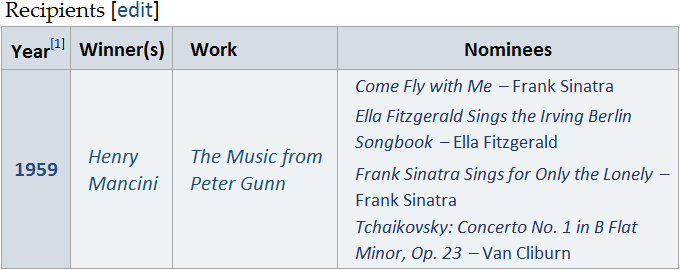}
        \caption{Example of Wikipedia table row.} \label{wiki1}
\end{figure}

The process of creating RELink queries involves two steps: (1) automatic selection of tables and columns within tables and (2) manual specification of information needs. For example,  in the table ``Grammy Award for Album of the Year'' the columns ``winner'', ``work'' were automatically selected to serve as entity types in the E-R query (Figure \ref{wiki1}). The relationship among these entities is suggested by the title and we let a human annotator to formulate the query.   

\begin{figure}[t]
    \centering
        \centering
        \includegraphics[width= 1.0\columnwidth]{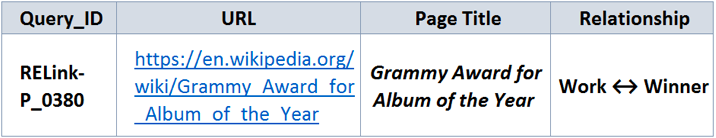}
        \caption{Example of metadata provided to editors.} \label{wiki2}
\end{figure}

The RELink query set was created by 6 annotators. We provided the annotators with access to the full table, metadata (e.g., table title or the first paragraph of the page) and entity pairs or triples to be used to specify the query (Figure \ref{wiki2}). For each entity pair or triple the annotators created a natural language information need and an E-R query in the relational format $Q = \{ Q^{E_i}, Q^{R_{i,j}}, Q^{E_j} \}$, as shown in Table \ref{qannot}. 
\subsection{Formulation of Queries}

The relational query format is introduced to support a variety of experiments with E-R queries. In essence, a complex information need is decomposed into a set of sub-queries that specify types of entities $E$ and types of relationships $R(E_i,E_j)$ between entities. For each relationship query there is one query for each entity involved in the relationship. Thus a query $Q$ that expects a pair of entities for a given relationship, is mapped into three queries $(Q^{E_i}$, $Q^{R_{i,j}}$, $Q^{E_j})$, where $Q^{E_i}$ and $Q^{E_j}$ are the entity types for $E_i$ and $E_j$ respectively, and $Q^{R_{i,j}}$ is a relationship type describing $R(E_i,E_j)$. For instance, ``football players who dated top models'' with answers such as $<$\textit{Cristiano Ronaldo}, \textit{Irina Shayk}$>$) is represented as three queries $Q^{E_i}=\{$\textit{football players}$\}$, $Q^{R_{i,j}}=\{$\textit{dated}$\}$, $Q^{E_j}=\{$\textit{top models}$\}$. Automatic mapping of $Q$ expressed in a natural language into queries $Q^{E_i}$ or $Q^{R_{i,j}}$ can be seen as a problem of query understanding \cite{yahya2012natural,pound2012interpreting,sawant2013learning} and is part of the future work. 
\begin{table}[t]
\caption{Examples of query annotations.}
\centering
\setlength\extrarowheight{3pt}
\begin{tabularx}{\columnwidth}{|l|X|X|}
\cline{1-3}
ID  & NL Query & Relational Format   \\ \cline{1-3}
\small RELink\_P\_164     & \small  \textit{What are the regiments held by the Indian Army?}   &   \small  \{\textit{regiment}, \textit{held by}, \textit{Indian Army}\}         \\ \cline{1-3}
\small RELink\_T\_071 &  \small  \textit{In which seasons NHL players scored more than 50 goals and the team they represented?}  &  \small    \{\textit{NHL season}, \textit{scored more than 50 goals in}, \textit{NHL player}, \textit{played for}, \textit{NHL team}   \}             \\ \cline{1-3}
\end{tabularx}
\label{qannot}
\end{table}

\begin{table}[t]
\caption{RELink collection statistics.}
\centering
\setlength\extrarowheight{1pt}
\begin{tabular}{l|l|l|l|}
\cline{2-4}
              & 2-entity  & 3-entity     & All   \\ \cline{1-4}  
\multicolumn{1}{|l|}{Total queries}      & 381  & 219 & \textbf{600} \\     \cline{1-4}
\multicolumn{1}{|l|}{Avg. queries length} & 56.5 & 83.8 & \textbf{66.5}   \\ \cline{1-4}
\multicolumn{1}{|l|}{Avg. $Q^E$ length} & 20.9 & 20.9 & \textbf{20.9}   \\ \cline{1-4}
\multicolumn{1}{|l|}{Avg. $Q^R$ length} & 11.8 & 12.6 & \textbf{12.3}   \\ \cline{1-4}
\multicolumn{1}{|l|}{\# uniq. entity attributes ($Q^E$)} & 679 & 592 & \textbf{1251}   \\ \cline{1-4}
\multicolumn{1}{|l|}{\# uniq. relationships ($Q^R$)} & 145 & 205 & \textbf{317}   \\ \cline{1-4}
\multicolumn{1}{|l|}{Avg. \# relevant judgments } & 67.9 & 41.8 & \textbf{58.5}   \\ \cline{1-4}

\end{tabular}
\label{stats}
\end{table}

\subsection{Collection Statistics}

RELink QC covers 9 thematic areas from the \textit{Lists-of-Lists-of-Lists} in Wikipedia: Mathematics and Logic, Religion and Belief Systems, Technology and Applied Sciences, Miscellaneous, People, Geography and Places, Natural and Physical Sciences, General Reference and Culture and the Arts. The most common thematic areas are Culture and the Arts with 70 queries and Geography and Places with 67 queries.

In Table \ref{stats} we show the characteristics of the natural language and relational queries. Among 600 E-R queries, 381 refer to entity pairs and 219 to entity triples. As expected, natural language descriptions of 3-entity queries are longer (on average 83.8 characters) compared to 2-entity queries (56.5 characters).

We further analyze the structure of relational queries and their components, i.e., entity queries $Q^E$ that specify the entity type and relationship queries  $Q^R$ that specify the relationship type.  Across 600 queries, there are 1251 unique entity types $Q^E$ (out of total 1419 occurrences). They are rather unique across queries: only 65 entity types occur in more than one E-R query and 44 occur in exactly 2 queries. The most commonly shared entity type is ``country'', present in 9 E-R queries.

In the case of relationships, there are 317 unique relationship types $Q^R$ (out of 817 occurrences) with a dominant type ``located in'' that occurs in 140 queries. This is not surprising since in many domains the key entity is tied to a location that is included in one of the columns. Nevertheless, there are only 44 relationship types $Q^R$ occurring more than once implying that RELink QC is a diverse set of queries, including 273 relationship types occurring only once. 

\section{Research Framework}

\begin{figure}[t]
    \centering
        \centering
        \includegraphics[width= 1.0\columnwidth]{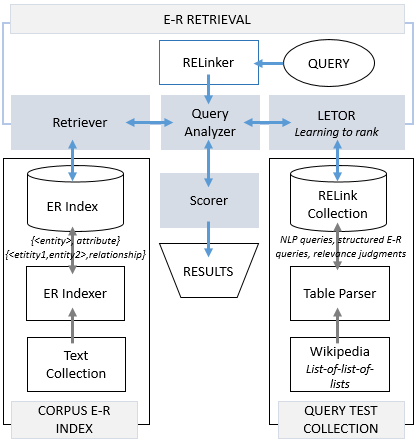}
        \caption{RELink Framework architecture overview.} \label{arch}
\end{figure}
RELink Framework is designed to facilitate experiments with the RELink QC. The RELink natural language queries and relational queries $(Q^{E_i}$, $Q^{R_{i,j}}$, $Q^{E_j})$ provide opportunities to define and explore a range of query formulations and search algorithms. A typical E-R experimental setup would involve search over a free-text collection to extract relevant instances of entity tuples and then verify their correctness against the relevance judgments derived from Wikipedia. The key enabling components therefore are: (1) test collections of documents with annotated entity instances that could be extracted during E-R search, (2) an indexing facility, and (3) a retrieval module to process queries and rank results. 

Currently, the RELink Framework includes the ClueWeb-09-B\footnote{\url{http://www.lemurproject.org/clueweb09/}} collection combined with FACC1\cite{gabrilovich2013facc1} text span annotations with links to Wikipedia entities (via Freebase). The entity linking precision and recall in FACC1 are estimated at 85\% and 70-85\%, respectively \cite{gabrilovich2013facc1}. The RELink Extractor, part of ER Indexer, applies an Open Information Extraction method \cite{schmitz2012open} over the annotated ClueWeb-09-B corpus. 
The two additional components are \textit{Corpus E-R Index} and \textit{E-R Retrieval}, both depicted in Figure \ref{arch}. The implementation of all modules in \textit{E-R Retrieval} and the Indexer module in \textit{Corpus E-R Index} are based on Apache Lucene and the Letor module serves as a wrapper for RankLib\footnote{\url{http://www.lemurproject.org/ranklib.php}}. 

\subsection{Indexing and Retrieval}
Based on the ClueWeb-09-B collection  we create two essential resources: \textit{entity index} and entity pair \textit{relationship index} for the entities that occur in the corpus. For a given entity instance, the ER Indexer identifies co-occuring terms within the same sentence and considers them as \textit{entity types} for the observed entity instance. Similarly, for a given pair of entities, the ER Indexer verifies whether they occur in the same sentence and extracts the separating string. That string is considered a context term for the entity pair that describes their \textit{relationship type}. We obtain 476M entity and 418M entity pair extractions with corresponding sentences that are processed by the Indexer. Once the inverted index (ER Index) is created, any instance of an entity or entity pair can be retrieved in response to the contextual terms, i.e., entity types and relationship types, specified by the users. 

\subsection{Search Process}
The  E-R retrieval process is managed by the RELinker module (Figure \ref{arch}). The Query Analyzer module processes information requests and passes queries in the structured format to the Retriever. Query search is performed in stages to allow for experimentation with different methods and parameter settings. First, the Retriever  provides an initial set of results using Lucene's default search settings and groups them by entity or entity pairs on query time using the Lucene's GroupingSearch. The Scorer then generates and applies feature functions of specific retrieval models with required statistics. Currently, the Scorer has implementations for Language Models \citep{elbassuoni2009language} and SDM \citep{metzler2005markov}. The RELinker is responsible for re-ranking and providing final results based on the scores provided by the Scorer and the parameter weights learned by Letor.
\section{Concluding remarks}
We anticipate that the RELink approach of using tabular data to create queries and relevance judgments will enable E-R research across different disciplines as researchers expand the RELink QC based on diverse sources of structured data. We recommend that the community retains the sources, e.g., tabular data used to create queries, in order to train methods for full automation of the query generation.

\begin{acks}
Authors would like to thank Jan \v{S}najder for assisting with acquiring and managing annotators of RELink queries.   
\end{acks}

\bibliographystyle{ACM-Reference-Format}
\bibliography{refs.bib} 


\begin{thebibliography}{00}


\ifx \showCODEN    \undefined \def \showCODEN     #1{\unskip}     \fi
\ifx \showDOI      \undefined \def \showDOI       #1{#1}\fi
\ifx \showISBNx    \undefined \def \showISBNx     #1{\unskip}     \fi
\ifx \showISBNxiii \undefined \def \showISBNxiii  #1{\unskip}     \fi
\ifx \showISSN     \undefined \def \showISSN      #1{\unskip}     \fi
\ifx \showLCCN     \undefined \def \showLCCN      #1{\unskip}     \fi
\ifx \shownote     \undefined \def \shownote      #1{#1}          \fi
\ifx \showarticletitle \undefined \def \showarticletitle #1{#1}   \fi
\ifx \showURL      \undefined \def \showURL       {\relax}        \fi
\providecommand\bibfield[2]{#2}
\providecommand\bibinfo[2]{#2}
\providecommand\natexlab[1]{#1}
\providecommand\showeprint[2][]{arXiv:#2}

\bibitem[\protect\citeauthoryear{Bhagavatula, Noraset, and Downey}{Bhagavatula
  et~al\mbox{.}}{2013}]%
        {bhagavatula2013methods}
\bibfield{author}{\bibinfo{person}{Chandra~Sekhar Bhagavatula},
  \bibinfo{person}{Thanapon Noraset}, {and} \bibinfo{person}{Doug Downey}.}
  \bibinfo{year}{2013}\natexlab{}.
\newblock \showarticletitle{Methods for exploring and mining tables on
  wikipedia}. In \bibinfo{booktitle}{{\em ACM SIGKDD Workshop on Interactive
  Data Exploration and Analytics}}. \bibinfo{pages}{18--26}.
\newblock


\bibitem[\protect\citeauthoryear{Conrad and Utt}{Conrad and Utt}{1994}]%
        {conrad1994system}
\bibfield{author}{\bibinfo{person}{Jack~G Conrad} {and}
  \bibinfo{person}{Mary~Hunter Utt}.} \bibinfo{year}{1994}\natexlab{}.
\newblock \showarticletitle{A system for discovering relationships by feature
  extraction from text databases}. In \bibinfo{booktitle}{{\em SIGIR\' 94}}.
  \bibinfo{pages}{260--270}.
\newblock


\bibitem[\protect\citeauthoryear{Elbassuoni, Ramanath, Schenkel, Sydow, and
  Weikum}{Elbassuoni et~al\mbox{.}}{2009}]%
        {elbassuoni2009language}
\bibfield{author}{\bibinfo{person}{Shady Elbassuoni}, \bibinfo{person}{Maya
  Ramanath}, \bibinfo{person}{Ralf Schenkel}, \bibinfo{person}{Marcin Sydow},
  {and} \bibinfo{person}{Gerhard Weikum}.} \bibinfo{year}{2009}\natexlab{}.
\newblock \showarticletitle{Language-model-based ranking for queries on
  RDF-graphs}. In \bibinfo{booktitle}{{\em CIKM}}. ACM,
  \bibinfo{pages}{977--986}.
\newblock


\bibitem[\protect\citeauthoryear{Gabrilovich, Ringgaard, and
  Subramanya}{Gabrilovich et~al\mbox{.}}{2013}]%
        {gabrilovich2013facc1}
\bibfield{author}{\bibinfo{person}{Evgeniy Gabrilovich},
  \bibinfo{person}{Michael Ringgaard}, {and} \bibinfo{person}{Amarnag
  Subramanya}.} \bibinfo{year}{2013}\natexlab{}.
\newblock \bibinfo{title}{FACC1: Freebase annotation of ClueWeb corpora}.
\newblock   (\bibinfo{year}{2013}).
\newblock


\bibitem[\protect\citeauthoryear{Lehmberg, Ritze, Meusel, and Bizer}{Lehmberg
  et~al\mbox{.}}{2016}]%
        {lehmberg2016large}
\bibfield{author}{\bibinfo{person}{Oliver Lehmberg}, \bibinfo{person}{Dominique
  Ritze}, \bibinfo{person}{Robert Meusel}, {and} \bibinfo{person}{Christian
  Bizer}.} \bibinfo{year}{2016}\natexlab{}.
\newblock \showarticletitle{A large public corpus of web tables containing time
  and context metadata}. In \bibinfo{booktitle}{{\em WWW}}.
  \bibinfo{pages}{75--76}.
\newblock


\bibitem[\protect\citeauthoryear{Li, Li, and Yu}{Li et~al\mbox{.}}{2012}]%
        {li2012entity}
\bibfield{author}{\bibinfo{person}{Xiaonan Li}, \bibinfo{person}{Chengkai Li},
  {and} \bibinfo{person}{Cong Yu}.} \bibinfo{year}{2012}\natexlab{}.
\newblock \showarticletitle{Entity-relationship queries over wikipedia}.
\newblock \bibinfo{journal}{{\em ACM TIST\/}} \bibinfo{volume}{3},
  \bibinfo{number}{4} (\bibinfo{year}{2012}), \bibinfo{pages}{70}.
\newblock


\bibitem[\protect\citeauthoryear{Metzler and Croft}{Metzler and Croft}{2005}]%
        {metzler2005markov}
\bibfield{author}{\bibinfo{person}{Donald Metzler} {and}
  \bibinfo{person}{W~Bruce Croft}.} \bibinfo{year}{2005}\natexlab{}.
\newblock \showarticletitle{A Markov random field model for term dependencies}.
  In \bibinfo{booktitle}{{\em SIGIR}}. ACM, \bibinfo{pages}{472--479}.
\newblock


\bibitem[\protect\citeauthoryear{Pound, Hudek, Ilyas, and Weddell}{Pound
  et~al\mbox{.}}{2012}]%
        {pound2012interpreting}
\bibfield{author}{\bibinfo{person}{Jeffrey Pound}, \bibinfo{person}{Alexander~K
  Hudek}, \bibinfo{person}{Ihab~F Ilyas}, {and} \bibinfo{person}{Grant
  Weddell}.} \bibinfo{year}{2012}\natexlab{}.
\newblock \showarticletitle{Interpreting keyword queries over web knowledge
  bases}. In \bibinfo{booktitle}{{\em CIKM}}. ACM, \bibinfo{pages}{305--314}.
\newblock


\bibitem[\protect\citeauthoryear{Sawant and Chakrabarti}{Sawant and
  Chakrabarti}{2013}]%
        {sawant2013learning}
\bibfield{author}{\bibinfo{person}{Uma Sawant} {and} \bibinfo{person}{Soumen
  Chakrabarti}.} \bibinfo{year}{2013}\natexlab{}.
\newblock \showarticletitle{Learning joint query interpretation and response
  ranking}. In \bibinfo{booktitle}{{\em WWW}}. ACM,
  \bibinfo{pages}{1099--1110}.
\newblock


\bibitem[\protect\citeauthoryear{Schmitz, Bart, Soderland, Etzioni,
  et~al\mbox{.}}{Schmitz et~al\mbox{.}}{2012}]%
        {schmitz2012open}
\bibfield{author}{\bibinfo{person}{Michael Schmitz}, \bibinfo{person}{Robert
  Bart}, \bibinfo{person}{Stephen Soderland}, \bibinfo{person}{Oren Etzioni},
  {et~al\mbox{.}}} \bibinfo{year}{2012}\natexlab{}.
\newblock \showarticletitle{Open language learning for information extraction}.
  In \bibinfo{booktitle}{{\em EMNLP-CoNLL}}. Association for Computational
  Linguistics, \bibinfo{pages}{523--534}.
\newblock


\bibitem[\protect\citeauthoryear{Yahya, Barbosa, Berberich, Wang, and
  Weikum}{Yahya et~al\mbox{.}}{2016}]%
        {yahya2016relationship}
\bibfield{author}{\bibinfo{person}{Mohamed Yahya}, \bibinfo{person}{Denilson
  Barbosa}, \bibinfo{person}{Klaus Berberich}, \bibinfo{person}{Qiuyue Wang},
  {and} \bibinfo{person}{Gerhard Weikum}.} \bibinfo{year}{2016}\natexlab{}.
\newblock \showarticletitle{Relationship queries on extended knowledge graphs}.
  In \bibinfo{booktitle}{{\em WSDM}}. ACM, \bibinfo{pages}{605--614}.
\newblock


\bibitem[\protect\citeauthoryear{Yahya, Berberich, Elbassuoni, Ramanath, Tresp,
  and Weikum}{Yahya et~al\mbox{.}}{2012}]%
        {yahya2012natural}
\bibfield{author}{\bibinfo{person}{Mohamed Yahya}, \bibinfo{person}{Klaus
  Berberich}, \bibinfo{person}{Shady Elbassuoni}, \bibinfo{person}{Maya
  Ramanath}, \bibinfo{person}{Volker Tresp}, {and} \bibinfo{person}{Gerhard
  Weikum}.} \bibinfo{year}{2012}\natexlab{}.
\newblock \showarticletitle{Natural language questions for the web of data}. In
  \bibinfo{booktitle}{{\em EMNLP-CoNLL}}. Association for Computational
  Linguistics, \bibinfo{pages}{379--390}.
\newblock


\end{thebibliography}

\end{document}